\documentclass[11pt]{article}

\usepackage{verbatim}
\usepackage{amsmath}
\usepackage{amsfonts}
\usepackage{amssymb}
\usepackage{bbm}
\usepackage{bbold}
\usepackage{latexsym}
\usepackage{lscape} 
\usepackage{hhline} 
\usepackage{epsfig}
\usepackage[usenames,dvipsnames]{color}
\usepackage{graphicx}
\usepackage{enumitem}
\usepackage[normalem]{ulem} 
\usepackage[table]{xcolor}
\usepackage{subfig}
\usepackage{multirow}
\usepackage{mathrsfs}
\usepackage{blkarray}
\usepackage{amscd}
\usepackage{cite}

\DeclareMathSymbol{\mg}{\mathrel}{symbols}{"1D}

\newcommand{\bes}{\begin{split}}
\newcommand{\ees}{\end{split}}
\renewcommand{\arraystretch}{1.5}

\def\beqn{\begin{eqnarray}}
\def\eeqn{\end{eqnarray}}

\usepackage{empheq}
\usepackage[most]{tcolorbox}
\usepackage{blkarray}

\newcommand{\cN}{{\cal N}}

\newcommand{\ra}{\rightarrow}

\newcommand{\beq}{\begin{equation}}
\newcommand{\eeq}{\end{equation}}
\newcommand{\barr}{\begin{array}}
\newcommand{\earr}{\end{array}}

\newcounter{oldcounter}

\newcommand{\Intr}{\mathbbm{Z}}

\newcommand{\ba}[2]{\[\begin{array}{#2}\label{#1}}
\newcommand{\ea}{\end{array}\]}
\newcommand{\be}{\begin{equation}}
\newcommand{\ee}{\end{equation}}
\newcommand{\bea}{\begin{eqnarray}}
\newcommand{\eea}{\end{eqnarray}}

\newcommand{\nn}{\nonumber}

\begin{document}

\thispagestyle{empty}

\begin{flushright}
LTH-1258\\ 
\end{flushright}
\vskip 1 cm
\begin{center}
{\Large {\bf 
Constraint on Spinor--Vector Dualities in Six Dimensions
} 
}
\\[0pt]

\bigskip
\bigskip {\large
{\bf A.E.~Faraggi$^{a,}$}\footnote{
E-mail: alon.faraggi@liverpool.ac.uk},
{\bf S.~Groot Nibbelink$^{b,c,}$}\footnote{
E-mail: s.groot.nibbelink@hr.nl},
{\bf   M.~Hurtado Heredia$^{a,}$}\footnote{
E-mail: martin.hurtado@liv.ac.uk}
\bigskip }\\[0pt]
\vspace{0.23cm}
${}^a$ {\it 
Department of Mathematical Sciences, University of Liverpool, Liverpool L69 7ZL, UK 
 \\[1ex] } 
${}^b$ {\it 
Institute of Engineering and Applied Sciences, Rotterdam University of Applied Sciences, \\ 
G.J.\ de Jonghweg 4 - 6, 3015 GG Rotterdam, the Netherlands
 \\[1ex]}
${}^c$ {\it 
Research Centre Innovations in Care, Rotterdam University of Applied Sciences, \\ 
Postbus 25035, 3001 HA Rotterdam, the Netherlands
 } 
\\[1ex] 
\bigskip
\end{center}

\subsection*{\centering Abstract}

Imprints of spinor--vector dualities have been uncovered in various string constructions. They are typically induced by changing certain free general GSO phases in the underlying string partition functions. This paper shows that spinor--vector dualities in six dimensions are constrained by a fundamental effective field theory consistency condition, namely that any six dimensional low energy theory must be free of irreducible $SO(2N)$ anomalies. Aspects of spinor--vector dualities are analysed in four six--dimensional free fermionic models which are distinguished by two generalised GSO phases. In addition, the constraint on the number of spinors and vectors is confirmed on generic spectra which may occur in K3 line bundle compactifications of the heterotic $E_8\times E_8$ string.

\newpage 
\setcounter{page}{1}

\section{Introduction}
\label{sc:Introduction}

Much of the string phenomenology research revolves around the relation between string theories and their effective low energy field theories. When starting with some effective supergravity theory the question is, whether it can be lifted to an ultra violet complete string theory or not; the latter is referred to as the ``Swampland Program'' (for review and references see {\it e.g.}~\cite{Palti:2019pca}) in contemporary nomenclature. An alternative approach is to investigate the effective field theory limit of exact string solutions. String theory provides a perturbatively consistent framework for the synthesis of gravity and the gauge interactions. Its internal consistency seems to mandate the existence of additional degrees of freedom beyond those observed in the Standard Models of particle physics and cosmology. The number of string theories in ten dimensions is relatively scarce, and includes five supersymmetric theories and eight that are not. Moreover, the supersymmetric versions, together with eleven dimensional supergravity, are believed to be different perturbative limits of a more fundamental theory, often dubbed $M$--theory. 

The extra dimensions are compactified on an internal space such that they are hidden from contemporary experiments. As the choice of compactifications may be constrained, but is certainly not fixed, by string theory, there exists a plethora of string theory vacua in lower dimensions. These vacua can be studied by using exact worldsheet constructions, as well as, effective field theory target space tools, that explore the low energy particle spectrum of string compactifications. Examples of worldsheet constructions are free fermionic models~\cite{Antoniadis1987a,AB} and toroidal orbifold compactifications~\cite{Dixon:1985jw,Dixon:1986jc}. These approaches are, in fact, closely related as particular fermionic formulations of the heterotic--string can be interpreted as $\Intr_2\times \Intr_2$ toroidal orbifold compactifications~\cite{Faraggi1994, Athanasopoulos:2016aws}. The study of Calabi--Yau compactifications with vector bundles~\cite{Candelas:1985en} utilises various effective field theory and cohomology methods. Using any of these approaches a strong effort has been made to construct low energy models which get surprisingly close to the Minimal Supersymmetric Standard Model: See refs.~\cite{Faraggi:1989ka, Faraggi:1991jr, Faraggi:1992fa, Cleaver1999,Faraggi:2006qa,Faraggi:2017cnh} for free fermionic constructions, refs.~\cite{Lebedev:2006kn,Lebedev:2007hv,Lebedev:2008un,Blaszczyk:2009in}
for orbifold realisations and refs.~\cite{Donagi:1999ez,Donagi:2000zs,Braun2005,Braun:2005nv,Bouchard:2005ag,Anderson:2009mh,Blaszczyk:2010db,Anderson:2011ns,Nibbelink:2015vha}
for smooth compactifications, respectively. 

Even though the space of low energy vacua of string theory is huge, there may exist symmetries that underlie the entire space of vacua in lower dimensions. Indeed, exact worldsheet string theories exhibit a rich symmetry structure which arises due to the exchange of massless and massive modes. Mirror symmetry may be considered as a prime example of symmetry between lower dimensional vacua~\cite{Greene:1990ud,Candelas:1989hd}. It was observed initially via the exact worldsheet CFT constructions with profound implications for the geometrical spaces that are utilised in the effective field theory limits~\cite{Candelas:1990rm}. It has been suggested that mirror symmetry is related to T--duality~\cite{Strominger:1996it}. In toroidal orbifold compactifications T--duality arises due to the exchange of the moduli of the internal six dimensional compactified manifold~\cite{giveon_94}. 

The fermionic $\Intr_2\times \Intr_2$ orbifold led to the observation of a new duality in the space of heterotic--string compactifications, dubbed {\em spinor--vector duality}: Models are related to each other by exchange of a number of spinorial plus anti--spinorial representations by the same number of vectorial representations of underlying $SO(12)$ or $SO(10)$ GUT groups in models with $\cN=2$ or $\cN=1$ spacetime supersymmetry,
respectively~\cite{Faraggi2007a, Faraggi2007b,Faraggi:2007ms,CatelinJullien:2008pc}.
The spinor--vector duality was initially observed by simple counting~\cite{Faraggi2007b}, using the classification tools developed in~\cite{Faraggi:2004rq,Faraggi2007a,Faraggi2007b} for the heterotic--string with unbroken $SO(10)$ GUT symmetry. These cases were shown to arise due to the exchange of certain discrete generalised GSO phases in the free fermionic formulation~\cite{ Faraggi2007b,CatelinJullien:2008pc}, or due to discrete torsions in a toroidal orbifold constructions~\cite{Angelantonj:2010zj,Faraggi:2011aw}. The spinor--vector duality observed in $\Intr_2\times \Intr_2$ orbifold compactifications generalises to exact string solutions with interacting internal CFT~\cite{Athanasopoulos:2014wha}. 

Just as mirror symmetry has profound implications for the internal manifolds in the effective field theory limits of string theories, the spinor--vector duality is anticipated to have similar such imprints in the space of heterotic string vacua. Furthermore, it suggests the existence of symmetries on the internal manifolds together with the vector bundles that correspond to the gauge degrees of freedom in the heterotic--string. It is imperative to understand how the modular properties that underlie the worldsheet string constructions constrain the effective field theory limits. At the very least, we expect the spinor--vector duality to provide a useful probe of the moduli spaces of the $(2,0)$ string compactifications.

\subsubsection*{Outline of our main finding}

This paper makes a modest step in this direction by exploring the realisation of the spinor--vector duality in compactifications of the heterotic--string to six dimensions. Even though this case provides a particularly controlled setting, because supersymmetry requires the two dimensional complex manifold to be essentially unique, albeit, realised in various ways both as $T^4/\Intr_K$ orbifolds or generic K3 geometries, it seems to have been omitted in the literature on the spinor--vector duality as far as we are aware. Our investigation of the spinor--vector duality makes use of both the string theory worldsheet tools as well as the effective field theory techniques. One central requirement on any effective field theory is that it is free of anomalies. Irreducible anomalies need to be absent entirely, while reducible anomalies may be compensated by some variant of the Green--Schwarz mechanism. If the effective field theory contains an $SO(2N)$ gauge group factor (immaterial of whether in the hidden or the observable sector), cancellation of its irreducible anomalies leads to a linear relation between the number of vectorial states and the sum of spinorial and conjugate spinorial representations constraining the possible realisations of any spinor--vector duality in six dimensions. Besides confirming this relation with many results in the literature, this relation is shown to be fulfilled in six dimensional models obtained from the free fermionic formulation with various choices of discrete generalised GSO phases. In addition, we confirm this result for any smooth K3 compactification with arbitrary line bundle gauge backgrounds.

\subsubsection*{Outline}

Our paper is organised as follows: To set the stage Section~\ref{sc:RecapSVD} provides a brief overview of the spinor--vector dualities in four dimensional models emphasizing that they are typically induced by changing generalised GSO phases. Section~\ref{sc:ConstraintNVNS} turns to six dimensional effective theories and establishes a linear relation between the number of vectorial and the sum of spinorial and anti--spinorial representations. Section~\ref{sc:FreeFermModels} considers a particular  free fermionic construction of models in six dimensional target space. Even though, both gauge groups and spectra strongly depend on the choices of two generalised GSO phases, the constraint on the number of $SO(2N)$ vectors and spinors is always respected. Finally, Section~\ref{sc:SmoothK3Models} considers generic smooth K3 realisations with line bundle gauge backgrounds. Using the six dimensional multiplicity operator, generic formulae to count the total number of vectorial and the number of (conjugate) spinor representations are both expressed in terms of the instanton number in the observable $E_8$. As expected, also in all these cases the linear relation on the number of vectorial and spinorial states is respected.


\section{Glancing overview of the spinor--vector duality}
\label{sc:RecapSVD}

This section gives a brief overview of some aspects of the spinor--vector duality in four dimensional $\Intr_2\times\Intr_2$ toroidal string vacua. The spinor--vector duality may be readily understood by considering the case in which an $SO(10)\times U(1)$ symmetry is enhanced to $E_6$. In this case the string compactification possesses an (2,2) worldsheet supersymmetry. The representation of $E_6$ are the chiral $27$ and anti--chiral $\overline{27}$, which decompose as
\begin{eqnarray}
27& = & 16_{+1/2}+10_{-1}+1_{+2}\nonumber
\\[1ex] 
\overline{27}& = &\overline{16}_{-1/2}+10_{+1}+1_{-2} \nonumber
\end{eqnarray}
under $SO(10)\times U(1)$. If one now counts the sum of the number of all $(16)$-- and $(\overline{16})$--plets  $N_S$ and total number of $(10)$--plets $N_V$, it is obvious that in this case $N_S = N_V$. Thus, the point in the moduli space in which the symmetry is enhanced to $E_6$, is a self--dual point under the spinor--vector duality. This is similar to the case of T--duality on a circle, in which at the self--dual radius under T--duality the gauge symmetry is enhanced from $U(1)$ to $SU(2)$.

Away from the self--dual point the $E_6$ symmetry is broken to $SO(10)\times U(1)$ and the worldsheet supersymmetry is broken from $(2,2)$ to $(2,0)$. In general the $E_6$ symmetry is broken in the toroidal orbifold models by Wilson lines, or by some discrete phases, whereas in the fermionic language they may appear as generalised GSO phases in the one--loop partition function. The spinor--vector duality states that for any string vacuum, in which $E_6\rightarrow SO(10)\times U(1)$, with $N_S$ $(16)$-- plus $(\overline{16})$--plets and $N_V$ $(10)$--plets, there exist a dual vacuum in which $N_V \leftrightarrow N_S$. This may be realised concretely  as a map between two Wilson line backgrounds in the two dual models. Moreover, as the free fermionic heterotic--string vacua correspond to the $\Intr_2\times \Intr_2$ orbifolds, they contain three twisted sectors each preserving an $\cN=2$ spacetime supersymmetry. The spinor--vector duality can then be realised in each twisted sector separately, {\it i.e.}\ it can be realised
in models that possess $\cN=2$, rather than $\cN=1$, spacetime supersymmetry~\cite{Faraggi:2007ms}. In the $\cN=2$ vacua the enhanced symmetry at the self--dual point is $E_7$, which is broken to $SO(12)\times SU(2)$ away from the self--dual point, and the spinor--vector duality is realised in terms of the relevant representations of $E_7$ and $SO(12)\times SU(2)$ \cite{Faraggi:2007ms}.

Further insight into the structure underlying the spinor--vector duality is revealed by breaking the untwisted NS symmetry from $SO(16)\times SO(16)$ to $SO(8)^4$ and generating the $SO(10)$ or $SO(12)$ symmetries by enhancements~\cite{Faraggi:2007ms}. This is obtained by defining four modular basis vectors for the free fermionic formulation with four non--overlapping sets of four periodic fermions. In this case a $\Intr_2$ twist that acts in the internal dimensions breaks one $SO(8)$ symmetry to $SO(4)^2$. The modular basis vector which is charged under this $SO(8)$ acts as a spectral flow operator, in a similar way to the spectral flow operator that mixes between the spacetime supersymmetric multiplets on the supersymmetric side of the heterotic string. In the (2,2) vacua with enhanced $E_6$ (or $E_7$) symmetry, this modular basis vector acts as a symmetry generator of the enhanced symmetry and exchanges between the $SO(10)\times U(1)$ components inside the $E_6$ representations. When the $E_6$ symmetry is broken to $SO(10)\times U(1)$, this spectral flow operator induces the map between the dual Wilson lines and hence between the two spinor--vector dual vacua~\cite{Faraggi:2007ms,Faraggi:2011aw}.

In a bosonic representation of the spinor--vector duality~\cite{Angelantonj:2010zj,Faraggi:2011aw} the map between the dual vacua results from an exchange of a discrete torsion. For one choice of the discrete torsion, the zero modes of the untwisted torus in the $\cN=2$ twisted sector are attached to the spinorial characters of the GUT group, whereas for the other choice they are attached to the
vectorial character. The choice of discrete torsion can be represented as the choice of the Wilson line background. In the vacua possessing $\cN=2$ spacetime supersymmetry the mapping between the dual Wilson lines is continuous~\cite{Faraggi:2011aw}, whereas in those possessing $\cN=1$ it is discrete, as the moduli which allowed the continuous interpolation in the $\cN=2$ case are projected out.

\section{Constraint on the number of spinorial and vectorial states}
\label{sc:ConstraintNVNS}

This section shows that in any six dimensional $\cN=1$ supersymmetric effective field theory with the numbers of vectors $N_V$ and of spinors $N_S$ (of either chirality) of some $SO(2N)$ gauge group are constrained by an anomaly condition to 
\begin{equation} \label{eq:ConstraintNVNS} 
N_V  =  2^{N-5}\, N_S + 2N -8~, 
\end{equation}
for $N\geq 3$\,. This result is derived under the assumption that the effective field theory in six dimensions possesses at least $\cN=1$ supersymmetry (i.e.\ $\cN=2$ supersymmetry in four dimensions) and the only $SO(2N)$ charged states in the spectrum are hyper multiplets in the vector and spinor representations and a gauge multiplet in the adjoint. In particular, the effective theory may contain other gauge interactions with matter in arbitrary representations. If a hyper multiplet in the vector or spinor representations is also charged under other gauge groups, then the dimension of these representations are contained in the numbers $N_V$ and $N_S$\,. The validity of \eqref{eq:ConstraintNVNS}  can be checked for many six dimensional models in the literature. For example, all the perturbative and non--perturbative six dimensional orbifold and line bundle resolution models mentioned in \cite{Aldazabal:1997wi} and \cite{Honecker:2006qz} all fulfil this equation.

\subsection{Irreducible $\boldsymbol{SO(2N)}$ anomaly in six dimensions}

To derive the equation~\eqref{eq:ConstraintNVNS}, recall that gauge and gravitational anomalies in 6D are dictated by anomaly polynomials $I_{8}$ eight-forms~\cite{Green:1984bx,Erler:1993zy}. For charged fermions the anomaly polynomial takes the form: 
\begin{equation}
I_{8|R} = \widehat{A}(R_2) \, \text{ch}_{R}(F_2) \Big|_8~, 
\end{equation}
where the rooth--genus $\widehat{A}(R_2)$ as a function of the curvature two--form $R_2$ encodes gravitational anomalies and the Chern character 
\begin{equation}
\text{ch}_{R}(F_2) = 
\text{tr}_R \Big[ e^{i\tfrac{F_2}{2\pi}}\Big]~, 
\end{equation}
depends on the field strength two--form $F_2$ of the gauge theory and on the representation $R$ the fermions are in. For an effective theory to be consistent, all irreducible gauge and gravitational anomalies have to cancel among themselves. (Reducible anomalies may be cancelled by some variant of the Green--Schwarz mechanism~\cite{Green:1984sg,Green:1984bx}.) This goes in particular for irreducible $SO(2N)$ anomalies which is the sole focus of this section.

In six dimensions irreducible $SO(2N)$ anomalies, proportional to $\text{tr}_V(iF_2)^4$ (where the trace is over the vector representation of $SO(2N)$), are possible and therefore their sum need to vanish. Derivations of relevant trace identities are recalled in Appendix~\ref{sc:TraceIdentities}. In light of the assumptions on the effective six dimensional supersymmetric theories under investigation, there are only three contributions to the irreducible $SO(2N)$ anomalies to be considered: 
\begin{enumerate}
\item $N_V$ Hyper multiplets in the vector representation:
\begin{equation}
I_{8|V} \supset N_V \, \frac 1{4!} \text{tr}_V \Big(i \dfrac{F_2}{2\pi} \Big)^4
\end{equation}
This is obtained directly by expanding the Chern character to fourth order. Here, $\supset$ indicates that only irreducible $SO(2N)$ anomalies are considered, ignoring gravitational, other gauge and (mixed) reducible anomalies. 
\item Gauge multiplet in the adjoint representation:
\begin{equation}
I_{8|Ad} \supset - \frac 1{4!} \Big[
(2N-8)\, \text{tr}_V \Big(i \dfrac{F_2}{2\pi} \Big)^4
+ 3\, \Big( \text{tr}_V \Big(i \dfrac{F_2}{2\pi} \Big)^2 \Big)^2
\Big]
\supset -(2N-8)\, \frac 1{4!} 
 \text{tr}_V \Big(i \dfrac{F_2}{2\pi} \Big)^4~. 
\end{equation}
This result is derived in~\eqref{eq:AdTrace}. The minus sign out front is due to the fact that the gauginos in six dimensions have the opposite chirality as the hyperinos. 
\item $N_S$ Hyper multiplets in the (conjugage) spinor representation:
\begin{equation}
I_{8|S} \supset  N_S \,\frac 1{4!}\, 2^{N-5}\Big[ 
- \text{tr}_V \Big(i \dfrac{F_2}{2\pi} \Big)^4
+ \dfrac{3}{4}\, \Big( \text{tr}_V \Big(i \dfrac{F_2}{2\pi} \Big)^2 \Big)^2
\Big]
\supset -2^{N-5}\, N_S\, \frac 1{4!} 
 \text{tr}_V \Big(i \dfrac{F_2}{2\pi} \Big)^4~. 
\end{equation}
A derivation of this result can be found in~\eqref{eq:SpinorTrace}. 
\end{enumerate}
The total irreducible $SO(2N)$ anomaly is the sum of these three contributions. It only vanishes if the sum of their pre--factors do, which is precisely condition~\eqref{eq:ConstraintNVNS}.

\section{Six Dimensional Free Fermionic Models}
\label{sc:FreeFermModels}

\newcommand{\cc}[2]{c{#1\atopwithdelims[]#2}}
\newcommand{\mathleft}{\@fleqntrue\@mathmargin0pt}
\newcommand{\mathcenter}{\@fleqnfalse}

\subsection{Generalities of free fermionic description}

The six dimensional heterotic string is described  in terms of $16$ left--moving  and $40$ right--moving two dimensional real fermions in the free fermionic formulation in the light--cone gauge~\cite{Antoniadis1987a, AB, Kawai1987}. The string models are defined by specifying different phases picked up by fermions ($f_A, A=1,\dots,56$) when transported along the non--trivial cycles of the vacuum--to--vacuum amplitude. Each model corresponds to a particular choice of fermion phases consistent with modular invariance
that can be generated by a set of $N_B$ basis vectors $B = \{ v_a, a=1,\dots, N_B \}$, where each 
\[
v_a=\left\{\alpha_a(f_1),\ldots,\alpha_a(f_{56}))\right\}~,
\]
dictates the transformation properties of each fermion
\begin{equation}
f_A\to -e^{i\pi\alpha_a(f_A)}\ f_A~,  
\end{equation}
$A=1,\ldots, 56$\,. It is important to emphasise that the free fermionic formalism is identical to the free bosonic formalism, {\it i.e.}\ to toroidal orbifold compactifications, which follows from the equivalence of bosons and fermions in two dimensions. The two formulations are therefore describing the same physical object, albeit using different language and tools. While detailed dictionaries  exist~\cite{Athanasopoulos:2016aws}, translating a model from one representation to another can often be non--trivial. Each formalism carries its advantages and are in that respect complementary. In the free fermionic formalism all the moduli are taken a priori on equal footing and there is minimal structure, or not at all, to begin with. This has the advantage that some discrete torsions, for which some implicit choice has been made in orbifold constructions, are revealed at a very basic level in fermionic models. On the other hand, in the toroidal orbifold models there is a clearer distinction between the internal and the Wilson line moduli facilitating making contact with the smooth effective field theory limits in this approach.

The basis vectors generate a space $\Xi$ which contains $2^{N_B}$ sectors that produce the string spectrum. Each sector arises as a combination of the basis vectors
\begin{equation}
\beta = \sum n_a v_a,\ \  n_a =0,1~. 
\end{equation}
The spectrum is truncated by generalised GSO projections whose actions on a
string state $|\text{state}\rangle_\beta$ are given by
\begin{equation}\label{eq:gso}
e^{i\pi v_a\cdot F} |\text{state}\rangle_\beta = \delta_{\beta}\ \cc{\beta}{v_a}^* |\text{state}\rangle_\beta~,
\end{equation}
where $\vert\text{state}\rangle_\beta$ is a state with the vacuum defined by the worldsheet fermions that are periodic in the sector $\beta$ with possibly fermionic oscillators acting on it, which are counted by $F$, the fermion number operator, and $\delta_{\beta}=\pm1$ is the spacetime spin statistics index. Different choices for unfixed generalised GSO phases affect the states that remain massless in each of the sectors containing these basis vectors. In particular, some vectors act as projectors on some states, and do not merely fix some $U(1)$ charges of periodic fermions, when there is no overlap of periodic fermions between the basis vectors. Hence, different sets of projection coefficients $\cc{\beta}{v_a}=\pm1$ consistent with modular invariance give rise to different models. In summary: a model is defined uniquely by a set of basis vectors $v_a,a=1,\dots,N_B$ and a set of $2^{N_B(N_B-1)/2}$ independent projections coefficients $\cc{v_a}{v_b}, a>b$\,.

\subsection{Four free fermionic $\boldsymbol{T^4/\Intr_2}$ orbifold models}

After this general outline of constructions of six dimensional target space model, the focus is now on a collection of basis vectors $B$ which may be interpreted as $T^4/\Intr_2$ orbifolds. To facilitate we divide the  two dimensional free fermions in the light--cone gauge as follows: $\psi^{1,\dots,4}, \chi^i,y^i, \omega^i$ (real left--moving fermions) and
$\bar{y}^i,\bar{\omega}^i$ (real right--moving fermions) with $i=3,\dots,6$ labeling the real torus directions and  ${\bar\psi}^{1,\dots,6}$, $\bar{\eta}^{2,3}$, $\bar{\phi}^{1,\ldots,8}$ (complex right--moving fermions). The models under investigation are generated by a basis $B$ 
\[ 
B=\{v_1,v_2,\dots,v_{5}\}~,
\]
of $N_B=5$ basis vectors defined as:
\begin{eqnarray}
v_1=1&=&\{\psi^{1,\dots,4}, \chi^{3,\dots,6},y^{3,\dots,6}, \omega^{3,\dots,6}\,|\, \bar{y}^{3,\dots,6},\bar{\omega}^{3,\dots,6};  \bar{\psi}^{1,\dots,6}, \bar{\eta}^{2,3},\bar{\phi}^{1,\dots,8}\}~,\nn\\
v_2=S&=&\{\psi^{1,\dots,4},\chi^{3,\dots,6}\}~,\nn\\
v_{3}=z_1&=&\{\bar{\psi}^{1,\dots,6}, {\bar\eta}^{2,3}\}~,\label{basis} \\
v_{4}=z_2&=&\{\bar{\phi}^{1,\dots,8}\}~,\nn\\
v_{5}=b_1&=&\{\psi^{1,\ldots,4},y^{3,\ldots,6}\,|\,\bar{y}^{3,\ldots,6};\bar{\psi}^{1,\dots,6}\}~. \nn
\end{eqnarray}
The inclusion of the vector $1$ in the additive group is mandated by the modular invariance constraints. The vector $S$ is the spacetime supersymmetry generator. The basis vectors $z_1$ and $z_2$ identify the observable and hidden sectors, respectively. The vector $b_1$ corresponds to the $\Intr_2$ twist in the corresponding $T^4/\Intr_2$ toroidal orbifold model on the $SO(8)$ lattice. In addition, for later use we define the linear combination 
\beq
e = 1+S+z_1+z_2=\{y^{3,\ldots,6}, \omega^{3,\ldots,6}\,|\,
{\bar y}^{3,\ldots,6}, {\bar\omega}^{3,\ldots,6}\}~, 
\label{evector}
\eeq
which {\em e.g.}\ induces a map between the periodic worldsheet fermions
$\{y^{3,4,5,6}|{\bar y}^{3,4,5,6}\}
\rightarrow
\{\omega^{3,4,5,6}|{\bar\omega}^{3,4,5,6}\}$ when mapping $b_1$ to $b_1+e$\,.

The matrix of one--loop generalised GSO phases is given by
\begin{equation}
\cc{\alpha}{\beta}=
{\bordermatrix{
              \beta\,\vline\, \alpha\! \! &  1&  S &{z_1} &{z_2} & {b_1}\cr
             1& -1&~~1 & -1   &  -1  &  -1   \cr
             S&~~1&~~1 & -1   &  -1  & ~~1  \cr
         {z_1}& -1& -1 &~~1   & \pm1 &  -1   \cr
         {z_2}& -1& -1 &\pm1  & ~~1  &\pm1   \cr
         {b_1}& -1& -1 &~~1   &\pm1  &  -1   \cr}}~. 
\label{freephases6D}
\end{equation}
Up to changes of internal chiralities, there is a twofold freedom in the choice of the generalised GSO phases: $\cc{z_1}{z_2}=\cc{z_2}{z_1}=\pm1$ and $\cc{b_1}{z_2}=\cc{z_2}{b_1}=\pm1$\,. These free generalised GSO phases can be translated to discrete torsion phases in the bosonic formalism.

\subsection{Gauge groups}

The gauge symmetry generated in the vacuum that contains the $\{1,S\}$ basis vectors is $SO(40)$\,. This gauge symmetry is broken by the basis vectors $z_1$ and $z_2$ to $SO(8)\times SO(16)\times SO(16)$\,. The generalised GSO phase $\cc{z_1}{z_2}=\pm1$ enhances the gauge symmetry to $SO(8)\times E_8\times E_8$ for $\cc{z_1}{z_2}=+1$, whereas with $\cc{z_1}{z_2}=-1$ the vector bosons arising from the sectors $z_1$ and $z_2$ are projected out and the gauge symmetry remains $SO(8)\times SO(16)\times SO(16)$. The inclusion of the final basis vector $b_1$ breaks $\cN=2$ six dimensional supersymmetry to $\cN=1$: The chirality of the gauginos and hyperinos are left-- and right--handed corresponding to an even/odd  number zero modes $\psi^\mu_0$ on their string vacuum states, respectively. Furthermore the basis vector $b_1$ reduces the gauge symmetry generated by the NS--sector alone to 
\beq
SO(4)_{{\bar y}^{3,\ldots,6}}\times
SO(4)_{{\bar\omega}^{3,\ldots,6}}\times  
SO(12)_{{\bar\psi}^{1,\ldots, 6}}\times
SO(4)_{{\bar\eta}^{2,3}}\times  
SO(16)_{{\bar\phi}^{1,\ldots, 8}},
\label{untwistedgg}
\eeq
where the subscripts indicate which worldsheet fermions generate the specified subgroup. As above if $\cc{z_1}{z_2}=-1$ no gauge enhancement occurs, while for $\cc{z_1}{z_2}=1$ the gauge group enhancement depends on the other generalised GSO phase $\cc{b_1}{z_2}$\,. The resulting gauge groups for the choices of the two free generalised GSO phases are summarised in the top two rows of Table~\ref{tb:GaugeSpectrumGGSO}.

\subsection{Hyper multiplet representations}

\begin{table}
\centering
\begin{tabular}{|c|c||c|c|c||c|c|c||c|}
      \hline
        \multicolumn{2}{|c||}{\text{Gen.\ GSO}}   & \multicolumn{3}{c||}{\text{Sectors $b_1 \oplus (b_1+z_1)$}}  & \multicolumn{3}{c||}{\text{$(b_1+e) \oplus (b_1+e+z_1)$}} &  $S+z_2$      
        \\ \hline\hline 
      $\cc{z_1}{z_2}$ & $\cc{b_1}{z_2}$ & ~~~$32_{\rm S}$~~~ & ~~~$12_{\rm V}$~~~ & ~~~$16_{\rm V}$~~~ & ~~~$32_{\rm S}$~~~ & ~~~$12_{\rm V}$~~~ & ~~~$16_{\rm V}$~~~ & $128_{\rm S}$ 
      \\ \hline\hline 
      + & + & out & out & in & out & out & in & in \\
      \hline
      + & - & in & in  & out & in & in  & out  & out \\
      \hline
      - & + & out  & in  & out & in  & out  & in & out \\
      \hline
      - & - & in  & out & in & out  & in & out & out\\
      \hline
    \end{tabular}
    \caption{The effect of the different choices of the generalised GSO phases $\cc{z_1}{z_2}$ and $\cc{b_1}{z_2}$ on the hyper spectrum of $SO(12)$ and $SO(16)$ spinors and vectors is displayed.}
    \label{eq:ProjectionsGGSO}
\end{table}

The choices of the free GGSO phases $\cc{z_1}{z_2}=\pm1$ and $\cc{b_1}{z_2}=\pm1$ affect, in particular, the states that remain in the spectrum as massless hyper multiplets. To illustrate this focus for example on spinorial and vectorial representations under the $SO(12)$ GUT group. The spinorial representations arise from the sectors $b_1$ and $b_1+e$, whereas the vectorial representations arise from the sectors $b_1+z_1$ and $b_1+e+z_1$. For example, the explicit generalised GSO projections in the $b_1$ and $b_1+z_1$ sectors are given by:
\begin{equation}
e^{i\pi z_2   \cdot F_{b_1}} |b_1\rangle = \delta_{b_1}\ \cc{b_1}{z_2} |b_1\rangle
\label{gsob1}
\end{equation}
and 
\beqn
e^{i\pi z_2   \cdot F_{b_1+z_1}}
     \left(
     \{{\bar\psi}^{1,\ldots,6},
     {\bar\phi}^{1,\ldots,8}\}|b_1+z_1\rangle\right) & = & \delta_{b_1+z_1}
     \cc{b_1+z_1}{z_2}
     \left(\{{\bar\psi}^{1,\ldots,6},
          {\bar\phi}^{1,\ldots,8}\}|b_1+z_1\rangle\right) \nonumber
          \\
          & = &  ~~-~\cc{b_1}{z_2}\cc{z_2}{z_1}
          \left(\{{\bar\psi}^{1,\ldots,6},
          {\bar\phi}^{1,\ldots,8}\}|b_1+z_1\rangle\right)~,\nonumber
\label{gsob1z1}
\eeqn
respectively, where the $\{~ \}$ brackets refer to the fermionic oscillators that act on the vacuum in the $b_1+z_1$ sector. As there is no overlap of periodic fermions between $b_1$ and $z_2$, it can be inferred from \eqref{eq:gso}, that the $z_2$ basis vector either projects out the spinorial states from the sector $b_1$ altogether or keeps them all in. Given that ${\bar\phi}^{1,\ldots,8}$ are periodic in $z_2$, whereas ${\bar\psi}^{1,\ldots,6}$ are anti--periodic, the $z_2$ basis vector selects the vectorial states from the sector $b_1+z_1$.  Similar projections operate in the sectors $b_1+e$ and $b_1+e+z_1$ using the $e$--vector \eqref{evector}. 

The spinorial and vectorial states for the different choices of $\cc{b_1}{z_2}$ and $\cc{z_1}{z_2}$ are displayed in Table~\ref{eq:ProjectionsGGSO}. We omitted there the multiplicities that arise from oscillators of the internal fermions $\{{\bar y}, {\bar\omega}\}^{3,\ldots, 6}$\,. Thus, we only list states that transform under the observable $SO(12)$ and hidden $SO(16)$ group factors. It is noted from this table that the spinor--vector duality is induced by the map
\beq
\cc{b_1}{z_2}\rightarrow -\cc{b_1}{z_2}~. 
\label{svdmap}
\eeq
The degeneracy with respect to the internal worldsheet fermions $\{y,\omega\,|\,{\bar y},{\bar\omega}\}$ is identical in the spinorial sectors $b_1,~b_1+e$ and the vectorial sectors $b_1+z_1,~b_1+e+z_1$. Hence, the counting with respect to the internal fermions is identical for the spinorial and vectorial representations. The sector $z_1$ induces the so--called $x$--map of refs.~\cite{Faraggi:1992yz,CatelinJullien:2008pc}.

\begin{table} 
\scalebox{0.9}{
\( \renewcommand{\arraystretch}{1.6} 
\begin{array}{|c||c|c|c|c|} 
\hline 
\big( \cc{z_1}{z_2}\,,~\cc{b_1}{z_2}\big) & \big(+1,+1\big) & \big(+1,-1\big) & \big(-1,+1\big) & \big(-1,-1\big) 
\\ \hline\hline  
\text{Gauge} & 
SO(4)\!\times\! SO(4) \times &
SO(4)\!\times\! SO(4) \times &
SO(4)\!\times\! SO(4) \times  & 
SO(4)\!\times\! SO(4) \times
\\[-1ex]
\text{Symmetry} & 
E_7 \times SU(2) \times SO(16) &
~~~E_7 \times SU(2) \times E_8~~~ &
SO(12) \!\!\times\!\! SO(4) \!\!\times\!\! SO(16) & 
SO(12)\!\!\times\!\! SO(4) \!\!\times\!\! SO(16) 

\\ \hline\hline 
\text{Sector} & \multicolumn{4}{c|}{\text{Hyper Multiplet Representations}}   
\\  \hline 
S & (4,4,1,1,1) & (4,4,1,1,1) & (4,4,1,1,1) & (4,4,1,1,1) \\[-1ex]  
 & &  & (1, 1, 12, 4, 1 ) & (1, 1, 12, 4, 1 ) 
\\ \hline
S\oplus (S+z_1) &  (1, 1, 56, 2, 1 )& (1, 1, 56, 2, 1 ) & & 
\\ \hline 
S+z_2 &  (1,1,1,1,128) & & & 
\\ \hline
b_1 &  &  & &(2_L,1,32,1,1) \\[-1ex]  
       &  &  & & (2_R,1,32,1,1)
\\ \hline
b_1 \oplus (b_1+z_1) & & (2_L,1,56,1,1) & & \\[-1ex] 
  & & (2_R,1,56,1,1) & & 
\\ \hline
b_1+z_1 & (2_L,1,1,2_L,16) & & (2_L,1,12,2,1) & (2_L,1,1,2_L,16) \\[-1ex]
               & (2_R,1,1,2_L,16) &  & (2_R,1,12,2,1) & (2_R,1,1,2_L,16) \\ 
               & & (2_L,4,1,2,1) & (2_L,4,1,2,1) & \\[-1ex]  
               & & (2_R,4,1,2,1) & (2_R,4,1,2,1) &                
\\  \hline
b_1+e &  &  & (1, 2_L,\overline{32},1,1) &  \\[-1ex]
           &  & & (1, 2_R,\overline{32},1,1) & 
\\   \hline  
(b_1 + e) \, \oplus & &(1, 2_L,56,1,1) & & \\[-1ex] 
(b_1+e+z_1) & & (1, 2_R,56,1,1) & & 
\\ \hline
b_1+e+z_1 & (1, 2_L,1,2_L,16) &  & (1, 2_L,1,2_R,16) & (1, 2_L,12,2_L,1) \\[-1ex]
                  & (1, 2_R,1,2_L,16) & & (1, 2_R,1,2_R,16) &  (1, 2_R,12,2_L,1) \\ 
               & & (4,2_L,1,2,1) & & (4,2_L,1,2_R,1) \\[-1ex]  
               & & (4,2_R,1,2,1) & & (4,2_R,1,2_R,1)                 
\\ \hline\hline 
SO(12) &  \text{Self--dual by} & \text{Self--dual by} & N_V =12 & N_V=12 \\[-1ex] 
N_V=2 N_S + 4 & \text{$E_7$ enhancement} &  \text{$E_7$ enhancement} & N_S=4 & N_S=4
\\ \hline 
SO(16) & N_V =16 & & N_V = 8 & N_V=8 \\[-1ex]  
N_V=8 N_S + 8 & N_S =1 & & N_S=0 & N_S=0 \\ 
\hline 
\end{array}
\)}
\caption{ \label{tb:GaugeSpectrumGGSO} 
The six dimensional gauge group and massless matter depend the choices of the free generalised GSO phases: $\cc{z_1}{z_2}$ and $\cc{b_1}{z_2}$\,. Only the sectors are indicated that lead to non--vanishing hyperino states to form hyper multiplets in target space. 
}
\end{table}

Table~\ref{tb:GaugeSpectrumGGSO} summarises the complete massless spectrum in the four models that arise due to the four choices of the phases $\cc{z_1}{z_2}$ and $\cc{b_1}{z_2}$. The sector $S$ gives rise to hyperinos in the $(1,1,12, 4, 1)$ representation of the gauge symmetry in~\eqref{untwistedgg} generated by the NS--sector alone. Since, when  $\cc{z_1}{z_2}=+1$ the $z_1$ sectors enhance the observable gauge symmetry to $E_7$, states in $S+z_1$ expand this representation to $(1,1,56,2,1)$ as can be seen in the second and third column of Table~\ref{tb:GaugeSpectrumGGSO}.

\subsection{Spinor--vector duality aspects}

A curious map between the two models occurs in the sector $z_2$. Indeed, when in addition $\cc{b_1}{z_2}=-1$, the left--moving oscillator acting on the vacuum in the physical states are $\psi^{1,\ldots,4}$, whereas when $\cc{b_1}{z_2}=+1$ they are $\chi^{3,\ldots,6}$. Therefore, only in the case $\cc{z_1}{z_2}=+1$, $\cc{b_1}{z_2}=-1$ the symmetry is enhanced from $SO(16)$ to $E_8$. Nevertheless, the total number of massless degrees of freedom is maintained. This is a manifestation of the phenomenon discussed in~\cite{Angelantonj:2010zj,Faraggi:2011aw} that under such maps, induced by the changes of discrete torsions in the partition function, the organisation of the number of degrees of freedom under the spacetime group factors may change, but the total number of massless degrees of freedom is preserved.

In the case $\cc{z_1}{z_2}=-1$  the gauge symmetry is not enhanced. The corresponding models in the final two columns exhibit the spinor--vector duality map, induced by the discrete change in~\eqref{svdmap}. Albeit these two cases are, in fact, self--dual under spinor--vector
duality, {\it i.e.}\ they contain an equal number of twisted spinorial and twisted vectorial representations of the $SO(12)$ GUT group, merely the $SO(12)$ chiralities of the spinorial states of these two models are opposite. 

The final rows of Table~\ref{tb:GaugeSpectrumGGSO} confirm condition~\eqref{eq:ConstraintNVNS} for these four six--dimensional free fermionic models. As can be inferred from this table, in all cases this condition is satisfied for both $SO(12)$ and $SO(16)$. The condition is automatically satisfied for $E_7$, as the $56\ra (32,1)+(2,12)$ is self--dual when branched to $SO(12)$ representations.

\section{Smooth K3 Line Bundle Models}
\label{sc:SmoothK3Models}

\subsection{K3 Geometries}

In four dimensions there is topologically just a single geometry that preserves supersymmetry, the so--called K3 surface. Particular geometrical descriptions of K3 can be obtained as orbifold resolutions~\cite{Nahm:1999ps}. In the discussion below such an interpretation is neither necessary nor implied. Depending on the set of independent divisors $\{D_\alpha\}$ labelled by $\alpha$ chosen the geometry may appear different. This work refers to divisors as two dimensional hyper surfaces as well as the associated two--forms interchangeably so that the context dictates how they should be interpreted. The intersection  numbers 
\begin{equation}
-\kappa_{\alpha \beta} = D_\alpha D_\beta = \int_{K3} D_\alpha D_\beta~, 
\end{equation}
may be determined by integrating over the K3 as a whole. The Euler number of K3
\begin{equation}
\int_{K3} c_2 = - \frac 12 \int_{K3} \text{tr} \Big(\dfrac{\mathcal{R}_2}{2\pi}\Big)^2 = 24
\end{equation}
is given by the integral over its second Chern class $c_2$\,. Here $\mathcal{R}_2$ is the anti--Hermitian holomorphic curvature two--form.

\subsection{K3 Line Bundles}

The $U(1)$ gauge background encoded as an anti--Hermitian Abelian gauge field strength two--form can be expanded in terms of the divisors as: \cite{Nibbelink:2007rd,Nibbelink:2007pn}
\begin{equation}
\dfrac{\mathcal{F}_2}{2\pi} = D_\alpha\, \mathsf{H}_\alpha~, 
\quad 
\mathsf{H}_\alpha =  V_\alpha^I\, \mathsf{H}_I~,
\end{equation}
where the sums over the basis divisors labelled by $\alpha$ and over the Cartan generators labelled by $I$ are implied. The Cartan generators $\mathsf{H}_I$ of $E_8\times E_8$ are normalized such that $\text{tr}\, \mathsf{H}_I \mathsf{H}_J = \delta_{IJ}$\,. The embedding of the line bundle background is therefore characterized by sixteen component line bundle (embedding) vectors $V_\alpha = (V_\alpha^I)$. (For translations to other characterizations see {\em e.g.}~\cite{Honecker:2006qz,Nibbelink:2015vha}.) Often it is convenient to split the line bundle vectors in contributions in the first and second $E_8$ as: $V_\alpha = (\vec{V}_\alpha)(\vec{V}_\alpha')$ where $\vec{V}_\alpha$ and $\vec{V}'_\alpha$ both have 8 entries. 

The fundamental consistency requirement of such backgrounds is determined from the integrated Bianchi identity $\text{tr}(\mathcal{F}_2^2) - \text{tr}(\mathcal{R}_2^2) = 0$\,. On K3 it can be cast in the form: 
\begin{equation}
\tfrac 12\, \kappa_{\alpha\beta}\, V_\alpha^I V_\beta^I = 24~. 
\end{equation}
Using the splitting of the line bundle vectors in contributions in both $E_8$'s this condition may be written as 
\begin{equation}
c+ c' = 24~, 
\qquad 
c = \tfrac 12\, \kappa_{\alpha\beta}\, \vec{V}'_\alpha \cdot \vec{V}'_\beta~, 
\quad 
c' = \tfrac 12\, \kappa_{\alpha\beta}\, \vec{V}'_\alpha \cdot \vec{V}'_\beta~, 
\end{equation}
introducing the instanton numbers $c$ and $c'$ in the first and second $E_8$. (If non--perturbative compactifications involving five--branes are considered the sum of the instanton numbers no longer need to add up to 24.) 

In six dimensions the full charged spectrum can be determined using the multiplicity operator $\mathsf{N}$ given by~\cite{Nibbelink:2007pn}:
\begin{equation}
\mathsf{N} = -\int_{K3} 
\Big\{ \dfrac 12\, \Big(\dfrac{\mathcal{F}}{2\pi}\Big)^2 - \dfrac 1{24}\, \text{tr} \Big( \dfrac{\mathcal{R}}{2\pi} \Big)^2 \Big\}
= \tfrac 12\, \kappa_{\alpha\beta}\, \mathsf{H}_\alpha \mathsf{H}_\beta - 2~. 
\end{equation}
This operator counts the number of fermions in a given representation and its sign determines the six dimensional chirality of the underlying fermionic states: It equals $-2$ on gaugino states, hence the multiplicity operator in six dimensions can be used to determine the gauge group unbroken by the line bundle background directly. The multiplicity operator $\mathsf{N}$ is positive on hyperinos as they have the opposite chirality in six dimensions as gauginos. Hence, if positive, it counts the number of hyper multiplets in a given representation of the gauge group.

\begin{table} 
\[ \renewcommand{\arraystretch}{1.6} 
\begin{array}{|c|c|c|c|c|} 
\hline 
E_8\ \text{Weight} & $SO(10)$\ \text{Repr.}  & \text{Multiplicity} 
\\ \hline\hline   
\pm(0,0,0, \underline{\pm1^2,0^3})(0^8) & (45) & -2  
\\ \hline 
\pm(1,0,0, \underline{\pm1,0^4})(0^8) & (10) & \displaystyle 
N_{(10)}^{~1} = \tfrac 12\, \kappa_{\alpha\beta}\, V_\alpha^1 V_\beta^1 -2 
\\  
\pm(0,1,0, \underline{\pm1,0^4})(0^8) &  &  \displaystyle 
N_{(10)}^{~2} = \tfrac 12\,\kappa_{\alpha\beta}\, V_\alpha^2V_\beta^2 -2 
\\  
\pm(0,0,1, \underline{\pm1,0^4})(0^8) &  &  \displaystyle 
N_{(10)}^{~3} = \tfrac 12\,\kappa_{\alpha\beta}\, V_\alpha^3V_\beta^3 -2 
\\  \hline
\pm ( \tfrac 12, \tfrac 12, \tfrac 12, \underline{-\tfrac 12^e, \tfrac 12^{5-e}})(0^8) & (16) & \displaystyle 
N_{(16)} = \tfrac 18\, \kappa_{\alpha\beta}\, (V_\alpha^1+V_\alpha^2+V_\alpha^3)(V_\beta^1+V_\beta^2+V_\beta^3)-2 
\\  \hline 
\pm ( -\tfrac 12, \tfrac 12, \tfrac 12, \underline{-\tfrac 12^o, \tfrac 12^{5-o}})(0^8) & (\overline{16}) & \displaystyle 
N_{(\overline{16})}^{~1} = \tfrac 18\, \kappa_{\alpha\beta}\, (-V_\alpha^1+V_\alpha^2+V_\alpha^3)(-V_\beta^1+V_\beta^2+V_\beta^3)-2 
\\ 
\pm ( \tfrac 12, -\tfrac 12, \tfrac 12, \underline{-\tfrac 12^o, \tfrac 12^{5-o}})(0^8) &  & \displaystyle 
N_{(\overline{16})}^{~2} = \tfrac 18\, \kappa_{\alpha\beta}\, (\phantom{-}V_\alpha^1-V_\alpha^2+V_\alpha^3)(\phantom{-}V_\beta^1-V_\beta^2+V_\beta^3)-2 
\\ 
\pm ( \tfrac 12, \tfrac 12, -\tfrac 12, \underline{-\tfrac 12^o, \tfrac 12^{5-o}})(0^8) &  & \displaystyle 
N_{(\overline{16})}^{~3} = \tfrac 18\, \kappa_{\alpha\beta}\, (\phantom{-}V_\alpha^1+V_\alpha^2-V_\alpha^3)(\phantom{-}V_\beta^1+V_\beta^2-V_\beta^3)-2 
\\ \hline  
\pm(0,1,\pm1,0^5)(0^8) & (1) & \displaystyle 
N_{(1)}^{\pm1} = \tfrac 12\,\kappa_{\alpha\beta}\, (V_\alpha^2 \pm V_\alpha^3)(V_\beta^2 \pm V_\beta^3) - 2 
\\  
\pm(1,0,\pm1,0^5)(0^8) & & \displaystyle 
N_{(1)}^{\pm2} = \tfrac 12\,\kappa_{\alpha\beta}\, (V_\alpha^1 \pm V_\alpha^3)(V_\beta^1 \pm V_\beta^3) - 2 
\\  
\pm(1,\pm1,0,0^5)(0^8) & & \displaystyle 
N_{(1)}^{\pm3} = \tfrac 12\,\kappa_{\alpha\beta}\, (V_\alpha^1 \pm V_\alpha^2)(V_\beta^1 \pm V_\beta^2) - 2 
\\  \hline 
\end{array}
\]
\caption{ \label{tb:SO(10)Spectrum} 
The SO(10) spectrum of gauge and hyper multiplet of a class of line bundle models. The hyper multiplet states are in the vector $(10)$, spinor $(16)$ or $(\overline{16})$ or singlet $(1)$ representations of SO(10) arising from a single $E_8$\,.}
\end{table}

\subsection{Counting the Number of SO(10) Vector, Spinor and Singlet States}

Consider line bundle vectors such that the first $E_8$ in the ten dimensional gauge group is generically broken to $SO(10)$\,: 
\begin{equation}
V_\alpha = (\vec{V}_\alpha)(\vec{V}'_\alpha)~, 
\qquad \vec{V}_\alpha = (V_\alpha^1,V_\alpha^2,V_\alpha^3,0^5)~.  
\end{equation}
Here it is assumed that the three entries, $V_\alpha^1$,  $V_\alpha^2$ and  $V_\alpha^3$, of these bundle vectors are sufficiently general that the unbroken gauge group indeed contains an $SO(10)$ factor which is not enhanced to a larger (exceptional) gauge group. By evaluating the multiplicity operator on all of the weights of the first $E_8$ leads to Table~\ref{tb:SO(10)Spectrum}. 

It follows from Table~\ref{tb:SO(10)Spectrum}, that the total number $N_V$ of $(10)$--plets only depends on the instanton number of the first $E_8$\,: 
\begin{equation}
N_{V} = N_{(10)}^{~1} + N_{(10)}^{~2} + N_{(10)}^{~3}  = \tfrac 12\,\kappa_{\alpha\beta}\, \vec{V}_\alpha \cdot \vec{V}_\beta - 6 = c -6~. 
\end{equation}
Similarly, the total number $N_S$ of $(16)$-- and $(\overline{16})$--plets also only depends on $c$\,: 
\begin{equation}
N_{S} =
N_{(16)} + N_{(\overline{16})}^{~1} + N_{(\overline{16})}^{~2} + N_{(\overline{16})}^{~3} 
= 
\tfrac 12\,\kappa_{\alpha\beta}\, \vec{V}_\alpha\cdot \vec{V}_\beta - 8 = c -8
\end{equation} 
Consequently, it is straightforward to verify, that \eqref{eq:ConstraintNVNS} is fulfilled, 
\begin{equation}
N_V - N_S = (c-6) - (c-8) = 2~,
\end{equation} 
for $N=5$\,. Notice that in this case, even for the total number of singlets $N_{(1)}$ coming from the fist $E_8$ one can obtain a similar result: 
\begin{equation}
N_{(1)} = N_{(1)}^{+1} + N_{(1)}^{-1} + N_{(1)}^{+2} + N_{(1)}^{-2} + N_{(1)}^{+3} + N_{(1)}^{-3}  
=
 2\,\kappa_{\alpha\beta}\, \vec{V}_\alpha\cdot \vec{V}_\beta - 12 = 4\, c - 12 
\end{equation} 
Hence the $SO(10)$ spectrum is completely fixed in terms of the instanton number $c$ of the first $E_8$. These results hold in particular, if one assumes that the line bundle vectors only have non--zero entries in the first $E_8$, i.e.\ $\vec{V}'_\alpha = 0$, and hence $c = 24$\,: 
\begin{equation}
N_{V} = 24 - 6 = 18~,  
\quad 
N_{S} =  24 - 8 = 16~, 
\quad 
N_V - N_S = 2~, 
\quad 
N_{(1)} = 4\cdot 24 - 12 = 84~. 
\end{equation} 

\subsection{Counting the Number of SO(2N) Vectors and Spinors}

This exercise can be generalised to line bundle vectors that have $1\leq n\leq 8$ non--zero entries in the observable $E_8$, i.e.\ $\vec{V}_\alpha = (V^1_\alpha, \ldots, V^n_\alpha)$ and the rest in the second $E_8$. The resulting gauge group is $SO(2N)$ with $N = 8-n$. 
The number of $SO(2N)$ vectors is given by: 
\begin{equation} \label{eq:NV_K3} 
N_{V} = \sum_{I=1}^n \Big( \tfrac 12\,\kappa_{\alpha\beta}\, \vec{V}_\alpha^I \vec{V}_\beta^I - 2\Big) = c -2n~. 
\end{equation}
The sum of the numbers of spinor and conjugate--spinor contribution can be computed as: 
\begin{equation} \label{eq:NS_K3}
N_{S} = \sum_{\vec{S}} \Big( \tfrac 12\,\kappa_{\alpha\beta}\, (\vec{S}\cdot\vec{V}_\alpha) (\vec{S}\cdot \vec{V}_\beta) - 2\Big) 
= 2^{n-1} 
\Big( \tfrac 12\,\kappa_{\alpha\beta}\, \dfrac{\vec{V}_\alpha}2 \cdot  \dfrac{\vec{V}_\beta}2- 2\Big) 
= 2^{n-3} (c -8)~,
\end{equation} 
since the set of spinor--configurations to be considered is $\big\{ \vec{S} = (\underline{-\tfrac 12^x, \tfrac 12^{n-x}})$ for $x \leq [n/2]\big\}$; $2^{n-1}$ in total. In the second step it is used that all the cross terms with different entries of the vectors $\vec{V}_\alpha$ and $\vec{V}_\beta$ cancel out, since all possible sign combinations are summed over, hence only ``squares'' $V_\alpha^I V_\beta^I$ remain. (This result can easily be verified directly for the cases $n=1,2$.) Notice that the expressions \eqref{eq:NV_K3} and  \eqref{eq:NS_K3} provide lower bounds on the instanton number: $c \geq 2n$ or $c \geq 8$, which ever is the strongest, otherwise the number of vectors or spinor become negative (which is impossible). 

The multiplicities of vectors \eqref{eq:NV_K3} and spinors  \eqref{eq:NS_K3} satisfy the condition~\eqref{eq:ConstraintNVNS} of vanishing irreducible $SO(2N)$ anomalies. Indeed, inserting these expressions in this condition leads to: 
\begin{equation}
N_V - 2^{N-5}\, N_S = c - 2n - 2^{N-5}\, 2^{n-3}(c-8) = 8 - 2n = 2N -8~, 
\end{equation}
using that $n=8-N$\,.

\section{Conclusion}
\label{sc:Conclusion}


Understanding the relationship between exact string theories and their effective field theory limit is of intense contemporary interest. The main effort is aimed at addressing the question when does an effective field theory of quantum gravity
have an embedding in string theory, and when it does not. The approach explored in this paper aimed to study this relationship in the opposite direction. Namely, the goal was to study the imprint of the spinor-vector duality, which was observed in exact string solutions, in the effective field theories resulting from four dimensional compactifications. 

The hallmark of string theories are their various duality symmetries, among them
mirror symmetry, which arises due to the interchange of complex and K\"ahler moduli of the internal six dimensional compactified manifold. In a specific realisation of mirror symmetry it was demonstrated due to exchange of a discrete torsion between the two orbifold twists of a $\Intr_2\times \Intr_2$ orbifold~\cite{Vafa:1994}. In addition to the moduli of the internal six dimensional
compactified manifolds, heterotic--string theories contain moduli that arise from the gauge degrees of freedom, {\it i.e.}\ the Wilson line moduli. Similar to the discrete torsion that exist between the orbifold twists, there are also discrete torsions associated with the Wilson line moduli. 

The spinor--vector duality operates by the exchange of the Wilson line moduli.
It is a remarkable transformation because seemingly from the point of view of the effective low energy field theory the dual vacua have entirely different physical content, {\it e.g.}\ one may contain a number of spinors of the underlying GUT group, whereas the other may contain the same number of vectors. However, from the point of view of string theory the two vacua are related by the spinor--vector duality transformation. The origin of the duality at the string level is clear. In the string theory massless and massive modes are exchanged in the dual vacua. Furthermore, the total number of degrees of freedom is preserved in the duality
transformations~\cite{Angelantonj:2010zj, Faraggi:2011aw}.

Mirror symmetry has profound implications on the mathematical properties of complex manifolds. Be it mirror symmetry, or the spinor--vector duality
espoused in this paper, the basic question being explored can be phrased as ``What is the imprint of the modular properties of the worldsheet realisation
of string vacua in their effective field theory limit?''. The string duality symmetries imply the existence of a structure in the space of string vacua, which is
obscured from the point of view of the effective field theory limits. Following the worldsheet description may provide guidance to unravel this web of hidden connections. 

In this paper we investigated the presence of spinor--vector dualities in the compactifications to six dimensions in the worldsheet description as well as in the effective field theory representation. These cases are particularly interesting to explore because of their rather restrictive nature. It is found that a spinor--vector duality also operate in these cases, albeit being self--dual under the spinor--vector duality map. The reason for this constraint could be traced to a fundamental consistency condition of any six dimensional effective field theory, namely that it needs to be free of irreducible gauge anomalies. 


\appendix 
\def\theequation{\thesection.\arabic{equation}} 

\section{SO(2N) Trace Identities}
\label{sc:TraceIdentities}

In order to make this paper self--contained this appendix derives a number of trace identities used in the anomaly analysis presented in Section~\ref{sc:ConstraintNVNS}. Most of the results presented here are known in the literature on anomalies, see {\em e.g.}~ \cite{Schellekens:1987xh,Erler:1993zy,vanRitbergen:1998pn} and textbooks like \cite{gsw_2,pol_2}.

\subsection{Traces in the vector representation}

In the vector representation, denoted by $V$, the Hermitian $SO(2N)$ gauge field strength two--form $F_2$ may be block diagonalized as: 
\begin{equation}
iF_2 = 
\left(
\begin{array}{cccccc}
 0 & F_2^1   &  & &   \\
-F_2^1  & 0  &  & &  \\
  &   & \ddots & & \\ 
 & & & 0 & F_2^N \\ 
 & & & - F_2^N & 0   
\end{array}
\right)~, 
\quad 
(iF_2)^2 = 
\left(
\begin{array}{cccccc}
 (F_2^1)^2 &  &  & &   \\
& (F_2^1)^2    &  & &  \\
  &   & \ddots & & \\ 
 & & & (F_2^N)^2 &   \\ 
 & & & & (F_2^N)^2   
\end{array}
\right)~, 
\end{equation} 
for certain two--form eigenvalues $F_2^I$ labeled by $I=1,\ldots, N$. Taking the trace over the vector representation of the $k$--th power of this it follows immediately: 
\begin{equation} \label{eq:VectorTrace} 
\text{tr}_V(iF_2)^{2k} = 2\, \sum_I (F_2^I)^{2k}~, 
\end{equation}
for $k\geq 1$; $\text{tr}_V(\mathbbm{1}) = 2N$\,.

\subsection{Traces in the adjoint representation}

The adjoint $Ad = [V]_2$ of $SO(2N)$ is the two times anti--symmetrisation of the vector representation $V$. In general, the Chern character of a two times anti--symmetrised representation $[R]_2$ is related to the Chern character of the representation $R$ via
\begin{equation}
\text{ch}_{[R]_2}(F_2) = \dfrac 12 \Big( \Big(\text{ch}_{R}(F_2)\Big)^2 - \text{ch}_R(2F_2) \Big)~. 
\end{equation}
By expanding this relation to fourth order leads to the identity  
\begin{equation} \label{eq:AdTrace}
\text{tr}_{Ad}(iF_2)^4 = 
(2N-8)\, \text{tr}_{V}(iF_2)^4 + 
3\, \Big(\text{tr}_{V} (iF_2)^2 \Big)^2 
~. 
\end{equation} 

\subsection{Traces in the chiral spinor representations}

To obtain trace identities for $SO(2N)$ spinor representations $S^\pm$ it is convenient to have an explicit basis for the spin--generators of $SO(2N)$. Like in the vector representation $V$, one may assume that the system is diagonalised and one is working on the Cartan of $SO(2N)$ generated by the following $N$ matrices 
\begin{equation}
 \Sigma_1  =   \tfrac 12 \sigma_3 \otimes \mathbbm{1}_2 \otimes \cdots \otimes \mathbbm{1}_2~,
 \quad  \ldots \quad 
 \Sigma_N =  \mathbbm{1}_2 \otimes \cdots \otimes \mathbbm{1}_2 \otimes \tfrac 12 \sigma_3~.
\end{equation}
obtained from $N$ times tensor products of the Pauli matrices $\sigma_i$. The $SO(2N)$ chirality operator $\widetilde{\Gamma}$ can be used to define projections on positive and negative chiral subspaces of the spinor representation 
\begin{equation}
P^\pm = \dfrac{\mathbbm{1}\pm\widetilde{\Gamma}}{2}~, 
\qquad 
\widetilde{\Gamma} = \sigma_3 \otimes \cdots \otimes \sigma_3~. 
\end{equation}
Expand the $SO(2N)$ gauge field strength in spinor representations in this Cartan basis can be expressed as 
\begin{equation} \label{eq:F2SpinorRepr} 
i F_2 = \sum F_2^I\, \Sigma_I~, 
\quad 
(i F_2)^2 = \dfrac 14\, \sum_I (F_2^I)^2 + \sum_{J\neq I} F_2^I F_2^J\, \Sigma_I \Sigma_J
\end{equation} 
for $N\geq 3$. (For $N=2$: $\Sigma_1 \Sigma_2 = \widetilde{\Gamma}/4$; for $N=1$ the second term does not exists. These case are ignored.) The traces over positive or negative chiral spinor representations $S^\pm$ are given by 
\begin{equation}
\text{tr}_{S^\pm} (iF_2)^{2k} = \text{tr}_S\Big[ (iF_2)^{2k} \dfrac{\mathbbm{1} \pm \widetilde{\Gamma}}{2} \Big]~, 
\qquad 
\text{tr}_{S^\pm} \mathbbm{1} = \text{tr}_S\Big[  \dfrac{\mathbbm{1} \pm \widetilde{\Gamma}}{2} \Big]
= 2^{N-1}~, 
\end{equation} 
over $2^N\times 2^N$ spinor matrices. 

The middle term of  the square of \eqref{eq:F2SpinorRepr},  
\begin{equation}
(i F_2)^4 = \Big(\dfrac 14\, \sum_I (F_2^I)^2\Big)^2 + 2 \Big(\dfrac 14\, \sum_I (F_2^I)^2\Big)  \sum_{J\neq I} F_2^I F_2^J\,\Sigma_I \Sigma_J 
 + \Big( \sum_{J\neq I} F_2^I F_2^J\,\Sigma_I \Sigma_J\Big)^2~, 
\end{equation}
vanish as the Pauli matrices are traceless. To evaluate the trace of the final term, note that 
\begin{equation}
\text{tr}_S \Big( \Sigma_I\Sigma_J \, \Sigma_K \Sigma_L \, \dfrac {\mathbbm{1}\pm \widetilde{\Gamma}}2 \Big) = 
2^{N-1}\, \big(\dfrac 14\big)^2 \Big( \delta_{IK}\delta_{JL} + \delta_{IL}\delta_{JK} \Big)~,  
\end{equation}
for $J\neq I$ and $K\neq L$. Thus, 
\begin{equation}
\text{tr}_S\Big[ 
\Big( \sum_{J\neq I} F_2^I F_2^J\,\Sigma_I \Sigma_J\Big)^2\,  
\dfrac {\mathbbm{1}\pm \widetilde{\Gamma}}2
\Big] 
= 
2^{N-4}\, \sum_{J\neq I} (F_2^I)^2 (F_2^J)^2
= 
2^{N-4}\,  \Big( \sum_I (F_2^I)^2 \Big)^2 - \sum_I (F_2^I)^4~. 
\end{equation}
Putting all contributions together leads to 
\begin{equation}
\text{tr}_{S^\pm} (iF_2)^{4} = 
2^{N-5} \Big\{ 
\Big( \sum_I (F_2^I)^2 \Big)^2 + 
2 \Big[\Big( \sum_I (F_2^I)^2 \Big)^2 - \sum_I (F_2^I)^4\Big]
\Big\}~.
\end{equation} 
Expression this in terms of traces over the vector representation finally leads to the identity  
\begin{equation} \label{eq:SpinorTrace} 
\text{tr}_{S^\pm} (iF_2)^{4} = 
2^{N-5} \Big\{ 
- \text{tr}_V(iF_2)^{4}  + \frac{3}4 \Big( \text{tr}_V(iF_2)^{2}  \Big)^2
\Big\}~.
\end{equation} 
This important result can also be found in~\cite{Erler:1993zy}.

\bibliographystyle{paper}
{\small
\bibliography{paper}
}
\end{document}